\documentclass[twoside]{ilcws08}
\usepackage[latin1]{inputenc}
\usepackage[dvips]{graphicx,epsfig,color}
\usepackage{wrapfig,rotating}
\usepackage{amssymb,amsmath,array}

\begin{document}
\title{
CALICE Si/W electromagnetic Calorimeter} 
\author{Marcel Reinhard on behalf of the CALICE Collaboration
\vspace{.3cm}\\
LLR - Ecole polytechnique, IN2P3/CNRS \\
Palaiseau - France
}

\maketitle
\begin{abstract}
The CALICE prototype for a Si/W electromagnetic calorimeter has been tested in large scale test beams. Several million events with electrons and hadrons of different energies and impact angles have been recorded. The energy resolution has been measured to be $\left( (16.6 \pm 0.1)/\sqrt{E(GeV)} \oplus 1.1 \pm 0.1 \right) \%$ with a linearity within the $1\%$ level. The next step will be the construction of a large scale prototype which will take realistic experimental costraints into account. This module will naturally benefit from the experience gained with the first prototype.
\end{abstract}
\section{Introduction: The Prototype}
The particle flow approach requires calorimeters with ultra-high granularity. The design of the Si/W ECAL has thus been based on this principle. It is composed of 30 layers, each layer having an array of 3 $\times$ 3 silicon matrices. These matrices or wafers are divided into 6 $\times$ 6 cells with an area of 1 $cm^2$. The mechanical structure consists of tungsten plates wrapped in carbon fiber, creating 15 alveolas into which so-called slabs are inserted. These slabs consist of one layer of tungsten between two layers of silicon wafers and their corresponding PCBs. The prototype is divided into three stacks, each consisting of 10 layers but with different thicknesses of the passive layers. The overall thickness is about 20 cm, corresponding to 24 $X_0$. A detailed description of all the technical and comissioning aspects can be found in a dedicated paper~\cite{AMpaper}. 
\section{Summary of the test beam efforts}
A large effort has been undertaken to complete a extensive test beam program comprising different beam types. Tab.~\ref{tab:tb} lists the different periods of data taking with particle beams. Every period included physics (hadrons, electrons) as well as calibration (muons) events at several different impact positions and angles from 0 to 45 degrees. Data taking over a period of three years allows also tests of the long-term stability of the detector and its components. For each period a detailed simulation model exists in the Mokka~\cite{MOKKA} framework representing the whole test beam setup, including parts of the beam delivery system, upstream detectors (if present) and the other participating CALICE calorimeters. 
\begin{table}
\centerline{\begin{tabular}{|l|c|c|r|}
\hline
Year & Location & Instrumented channels & Physics data taken \\
\hline
2006  & DESY & 5184 & $e^-$ from 1 to 6 GeV  \\
\hline
2006  & CERN & 6480 & $e^-$/$e^+$ : 6 - 45 GeV  \\
      &      &      & $\pi^-$/$\pi^+$ : 6 - 60 GeV \\
\hline
2007  & CERN & up to 9072 & $e^-$/$e^+$ : 6 - 90 GeV  \\
      &      &            & $\pi^-$/$\pi^+$\: 6 - 180 GeV \\
\hline
2008  & FNAL & 9720 & $e^-$/$e^+$ : 1 - 30 GeV  \\
      &      &      & $\pi^-$/$\pi^+$ : 1 - 60 GeV \\
\hline
\end{tabular}}
\caption{Summary of data taken in test beams.}
\label{tab:tb}
\end{table}
\section{Performance studies}
The studies presented here are based on the CERN 2006 data taking period if not stated explicitly. 
\subsection{Calibration}
The mean MIP response of a cell of the detector is about $45.5$ ADC counts. One can identify three different groups of cells with the same level of response which correspond either to different manufacturers or different production series of the same manufacturer. The long-term stability of the response has also been tested. The correlations between different sets of calibration constants obtained at different points in time are very strong. 
Only 9 of the 6480 cells that were equipped during the CERN 2006 running were showing no signal output, which corresponds to $0.14\%$. 
\subsection{Noise and pedestal}
A thorough analysis has been done checking the stability of the noise and pedestal values in time and their uniformity across the detector. The mean of the noise value over all channels for all runs taken in 2006 is about 5.9 ADC counts. At the MIP level, the detector shows thus a signal over noise ratio of about $7.75$ (i.e. $12.9\%$ of a MIP). A more detailed description of the analysis of noise, pedestal and calibration can be found in \cite{AMpaper}.
\subsection{Response to electrons}
The response of the ECAL prototype to electrons is subject to a dedicated paper~\cite{response_paper}. We will just outline the most important aspects here. 
\subsubsection{Event selection}
Five different cuts were used to clean the event samples. These cuts can be divided into two categories. The first one is used to reject background events, i.e. multi-particle events and residual pions in the electron beam. These cuts are based on a wide energy window, an upstream \v{C}erenkov counter when it was usable in the beam setup and a shower shape variable that is used to detect particles that started their showers before entering the ECAL. The second group are geometrical cuts. On the one hand to reject events with lateral leakage of showers close to the edges of the prototype and on the other to avoid energy losses in the inter-wafer gaps.
\subsubsection{Linearity and Resolution}
The conversion of the energy deposit from MIPs to GeV is done via the function $E_{mean} = \beta * E_{beam} - \alpha$ with $\beta = 266.5$ and $\alpha = 99.25$. The residuals to this linear function are within the $1\%$ level over the whole studied energy range while being consistent with zero. The calorimeter is thus consistent with zero non-linearity. A fit to the relative resolution of the measured energy as a function of the inverse square root of the beam energy, following the well-known parametrisation of a quadratic sum of a stochastic and a constant term leads to the following result:
\[Data: \frac{\Delta E_{meas}}{E_{meas}} = \left(\frac{16.6 \pm 0.1}{\sqrt{E(GeV)}} \oplus 1.1 \pm 0.1 \right) \% \]
\[MC: \frac{\Delta E_{meas}}{E_{meas}} = \left(\frac{17.3 \pm 0.1}{\sqrt{E(GeV)}} \oplus 0.5 \pm 0.1 \right) \% \]
This performance seems highly suitable for the particle flow approach.
\section{The next prototype - The EUDET Module}
\begin{figure}[hc]
\begin{minipage}[c]{0.45\columnwidth}
\centering
\includegraphics[scale=.33]{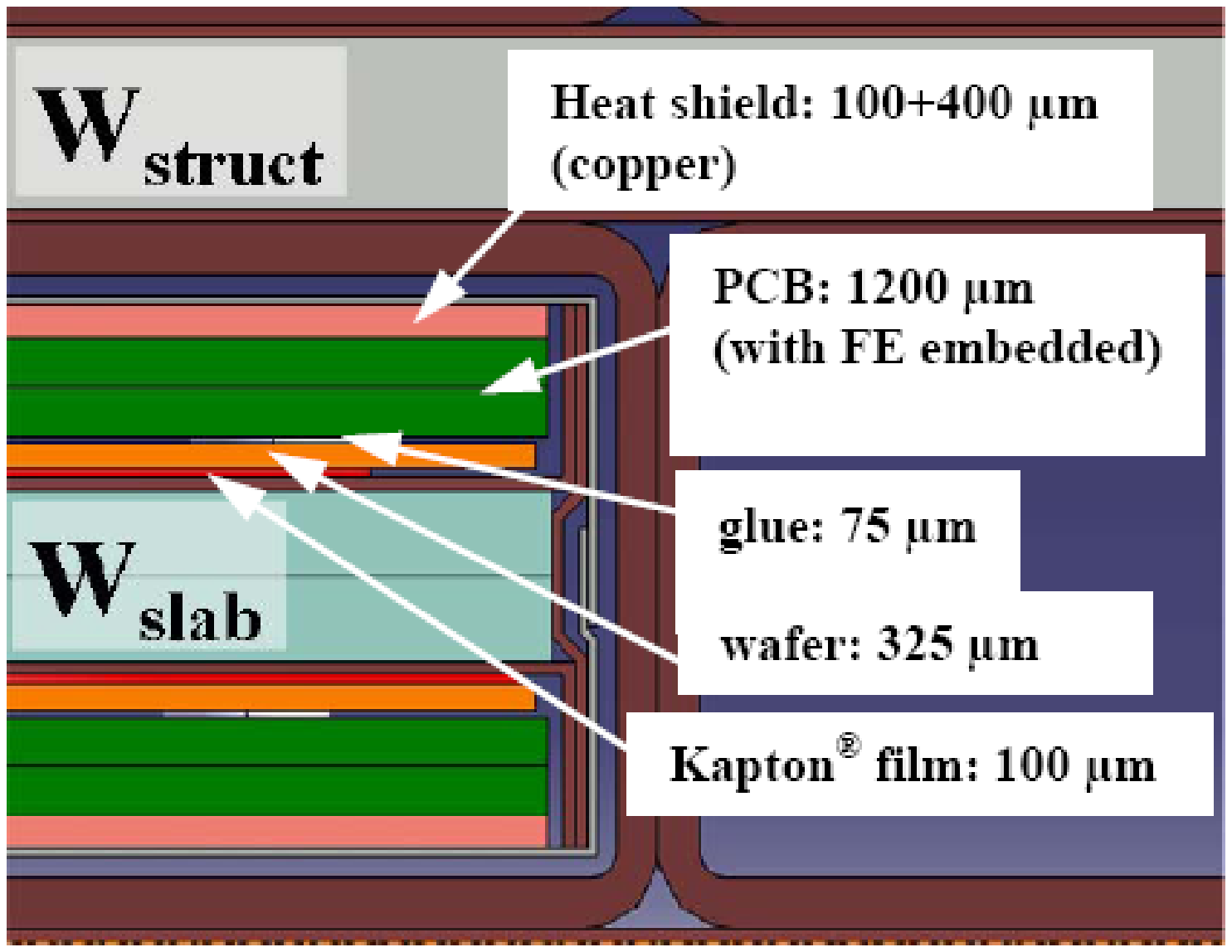}
\caption{Design for a detector slab of the EUDET module}
\label{Fig:EudetSlab}
\end{minipage}
\hspace{8mm}
\begin{minipage}[c]{0.45\columnwidth}
\centering
\includegraphics[scale=.2]{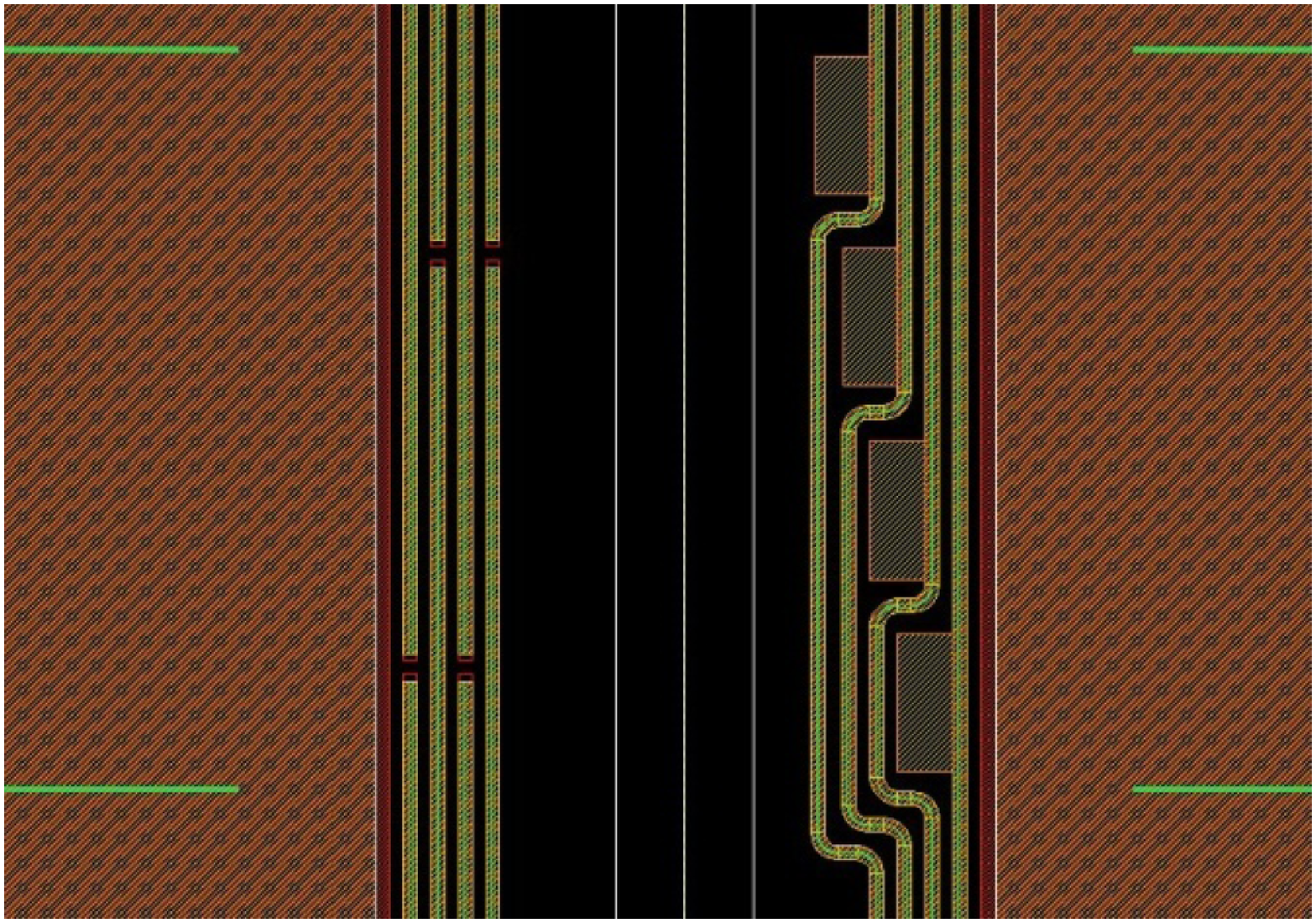}
\caption{Scheme of a possible solution to produce silicon wafers with segmented guardrings in order to supress the square event efffect}
\label{Fig:segmented_guardrings}
\end{minipage}
\end{figure}

While the physics prototype allowed to validate the basic concepts of the design of the Si/W ECAL, such as the alveolar structure, the use of detector slabs, gluing of the silicon wafers, the integration process and its physics capabilities, the next-generation prototype within the EUDET framework will concentrate on the study and validation of the technological solutions, i.e. integrated Very-Frontend Electronics, the cooling system and the feasability of producing large structures, thereby taking into account the industrialization aspect of the process to make a robust cost estimation per detector module. The actual evolution between the two prototypes towards a final detector module is illustrated in Table~\ref{tab:proto_dim}. More details can be found in \cite{EUDET}. A major point of the design is the compactness of the layers, as shown in Fiure~\ref{Fig:EudetSlab}). The total thickness foreseen for the Si wafers, the VFE and the cooling system is $2200\mu m$. Since the VFE will be power pulsed, the dissipated heat per channel will be as low as $25\mu W$. It is thus sufficient to drain the heat via copper sheets that are integrated in the slabs while only cooling actively at the extremities of a slab.
\begin{table}
\centerline{\begin{tabular}{|l|c|c|}
\hline
   & Physics prototype  & ILC-like prototype \\
\hline
$\#$ structures  & 3 ($10\times1.4mm +$  & 1 ($20\times2.1mm$ \\
(layers $\times$ W thickness) & $10\times2.8mm + 10\times4.2mm$) & $+ 10\times4.2mm$)\\
\hline
$X_0$ & 24 & $\sim23$ \\
\hline
Dimensions & $380 \times 380 \times 200 mm^3$ & $1560 \times 545 \times 186 mm^3$ \\
\hline
Slab thickness & $8.3mm$ ($W=1.4mm$) & $6mm$ ($W=2.1mm$) \\
\hline 
VFE & outside  & inside (zero-suppressed read-out) \\
\hline
$\#$ channels & 9720 & 45360 \\
\hline
Silicon wafer area (cellsize) & $6\times6 cm^2$ ($10\times10mm^2$) & $9\times9 cm^2$ ($5\times5mm^2$) \\
\hline
Weight & $\sim200kg$ & $\sim700kg$ \\
\hline
\end{tabular}}
\caption{Coparison of the Physics and ILC-like prototypes.}
\label{tab:proto_dim}
\end{table}
\subsection{Correction of encountered issues}
Several issues have been discovered during the tests with the physics prototype that will be corrected in the EUDET design. Some examples are given here: 

A fake differential in the PHY3 chip had as a consequence that the pedestal of a whole module is sensitive to exterior influences which led to instabilieties of the pedestal in time. The new chip design including a true differential will cure this problem. An event-by-event correction in the reconstruction proved to be efficient in the current data analysis. \\
It appeared that the pedestal in one wafer could drop with the strength of the total signal it recorded. A revision of the electrical circuit revealed a direct coupling to the bias voltage which caused this effect. A new desgin of the circuit will avoid this problem of signal induced pedestal shifts. 

Charge propagation over the external and internal guard rings of the silicon wafers led to a number of additional low-energy hits when a particle passes through one of these. A detailed description of this so-called square event effect and the inner wafer crosstalk can be found in \cite{AMpaper}. Significant R\&D work is ongoing to suppress this effect. One of the proposed solutions is to use segmented guard-rings to avoid propagation of the charge. A design example is shown in Fig~\ref{Fig:segmented_guardrings}. 

To minimize the ratio of the dead zone caused by the inter-wafer gaps and the active area, the next prototype will use silicon wafers with a larger area ($9\times9cm^2$ instead of $6\times6cm^2$ in the physics prototype). Minimizing the area occupied by the guard rings is another subject of ongoing studies.
\subsection{Demonstrator studies}
As an intermediate step before the EUDET module, a demonstrator module is being assembled. It will be used to validate additional techniques. This includes the composite processing, thermal studies with a specially prepared long PCB and cooling system, slab integration (with gluing and interconnection), possibilities for the attachement to the HCAL, etc. The dimensions of its frontface will be $1300\times380mm^2$ and it will include an alveolar structure with space for 3 slabs.
\section{Conclusions}
The data analysis on the current CALICE Si/W ECAL prototype has been fruitful. The measurement shows that the energy resolution of the physics prototype is as expected and is suitable for the particle flow approach. Also the agreement with the Monte Carlo simulation is quite satisfying, the amount of detail in the simulated model already being to such a high degree that problems can be easily identified. A number of studies, like longitudinal/transversal shower shapes, different impact angles, hadron analysis, clustering algorithms, gap correction and optimisation, cross-talk suppression and response stability (over $\sim3$years) are still ongoing. The intensive testing of this prototype led to a number of hardware modifications to improve the performance of the upcoming EUDET prototype. Its design has been fixed completely and the actual construction phase will start in 2009. In the meanwhile additional test are performed on a demonstrator module.


\begin{footnotesize}



%

\end{footnotesize}


\end{document}